\documentclass[letterpaper,14pt]{article}
\usepackage{setspace}
\doublespace
\usepackage{cite}
\usepackage{amsmath}
\usepackage{amsfonts}
\usepackage{amssymb,multirow,wrapfig}
\usepackage{color}
\usepackage{xfrac}
\usepackage{xcolor}
\usepackage{graphicx}
\usepackage{siunitx}
\colorlet{shadecolor}{blue!20}
\usepackage{framed}

\DeclareMathOperator{\sinc}{sinc}
\usepackage{epstopdf}

\usepackage[small,normal,up]{caption2}
\newcommand{\abs}[1]{\ensuremath{\left| #1 \right|}}

\newcommand{\be}[0]{\begin{equation}}
\newcommand{\ee}[0]{\end{equation}}
\newcommand{\bea}[0]{\begin{eqnarray}}
\newcommand{\eea}[0]{\end{eqnarray}}

\makeatletter
\g@addto@macro\normalsize{%
  \setlength\abovedisplayskip{2pt}
  \setlength\belowdisplayskip{2pt}
  \setlength\abovedisplayshortskip{2pt}
  \setlength\belowdisplayshortskip{2pt}
}
\makeatother

\begin{document}

%%%%%%%%%%%%%%%%%%%%%%%%%%%%%%%%%%%%%%%%%%%%%%%%%%%%%%%%%%%%
\setcounter{secnumdepth}{0}

\begin{center}

%% title -  Azimuthal Hanbury Brown and Twiss Interference and Angular-Position\textemdash Orbital-Angular-Momentum Correlations of Pseudothermal Light}
\LARGE\textbf{Hanbury Brown and Twiss Interferometry with Twisted Light}

%% authors
\vspace{4mm} \normalsize\textbf{Omar S. Maga\~{n}a-Loaiza$^{1,*}$, Mohammad Mirhosseini$^1$, Robert M. Cross$^1$, Seyed Mohammad Hashemi Rafsanjani$^1$,  and Robert W. Boyd$^{1,2}$}

%% affiliations
\vspace{1mm}
\footnotesize
\textit{$^1$The Institute of Optics, University of Rochester,
Rochester, New York 14627 USA}\\
\textit{$^2$Department of Physics and Max Planck Centre for Extreme and Quantum Photonics, University of Ottawa, Ottawa, ON K1N 6N5 Canada}\\
\textit{$^*$omar.maganaloaiza@rochester.edu}

\end{center}
\vspace{4mm}

\bigskip

\textbf{The rich physics exhibited by random optical wave fields permitted Hanbury Brown and Twiss to unveil fundamental aspects of light. Furthermore, it has been recognized that optical vortices are ubiquitous in random light and that the phase distribution around these optical singularities imprints a spectrum of orbital angular momentum onto a light field. We demonstrate that random fluctuations of intensity give rise to the formation of correlations in the orbital angular momentum components and angular positions of pseudothermal light. The presence of these correlations is manifested through distinct interference structures in the orbital angular momentum-mode distribution of random light. These novel forms of interference correspond to the azimuthal analog of the Hanbury Brown and Twiss effect. This family of effects can be of fundamental importance in applications where entanglement is not required and where correlations in angular position and orbital angular momentum suffice. We also suggest that the azimuthal Hanbury Brown and Twiss effect can be useful in the exploration of novel phenomena in other branches of physics and in astrophysics.}

\section{Introduction}
In 1956, Hanbury Brown and Twiss (HBT) revolutionized optical physics with the observation of a new form of interference produced by correlations of the intensity fluctuations of light from a chaotic source.   Their stellar interferometer collected light produced by independent sources on the disc of a star and detected at two different locations on Earth \cite{HBT1}. The observation of a second-order interference effect in this configuration was intriguing because at that time it appeared that classical and quantum theories of light offered different predictions \cite{Knight:2005cu}. Ever since, this effect has motivated extensive studies of higher-order classical correlations and their quantum counterparts in optics, as well as in condensed matter and particle physics \cite{Oppel:2014wp, Bromberg:2010jd, Baym:1997ce, Schellekens:2005cu}. Fundamental bounds have been established for the degree of correlation for a wide variety of degrees of freedom, such as in polarization, time, frequency, position, transverse momentum, angular position and orbital angular momentum (OAM) \cite{Pan:2012tm, Jha:2010gf, Leach:2010bm}.  

The random nature of light is an essential element of the HBT effect.  Moreover, the random properties of light have been investigated and applied in a wide variety of other contexts.  For example, speckled light, intimately related to pseudothermal light, has played a fundamental role in the development of  optical physics, imaging science, and nanophotonics.  In addition, the study of fundamental processes such as transport phenomena, localization of light, optical vortices, and optical correlations has led to the development of novel physics produced as a consequence of the chaotic properties of light \cite{DeRaedt:1989vd, Romero:2011dn, Xiong:2005kj, Gatti:2004ec, Nevet:2011dm, Zhang:2009kl}. These results have motivated interest in the design of random lasers and of disordered structures that scatter light in random directions, which serve as sources of pseudothermal light \cite{Redding:2012kd}.

As identified by Berry, optical vortices produced by the interference of random waves are intrinsic elements in chaotic light\cite{Berry:1978ga, Dennis:2009dj}. Interest in this field has exploded since the recognition of a special class of vortices that carry OAM, characterized by an azimuthal phase dependence of the form $e^{i\ell\phi}$, where $\ell$ is the OAM mode number and $\phi$ is the azimuthal angle \cite{Allen:1992vk}.  The azimuthal properties of light, described by the conjugate variables of angular position and OAM, have shown potential for technological applications in information science, remote sensing, imaging, and metrology \cite{Yao:2011uo}. In astrophysics, recent theoretical studies have predicted that rotating black holes can imprint an OAM spectrum onto  light. The measurement of this spectrum could lead to an experimental demonstration of the existence of rotating black holes \cite{Tamburini:2011wm}. In addition, the optical vortex coronagraph has allowed the observation of dim exoplanets by canceling a diffraction-limited image of a star \cite{Foo:2005iy}. More recently, it has been proposed to use rotational Doppler shifts for astronomy \cite{Lavery:2013iz}.

Here, we show that random fluctuations give rise to the formation of intensity correlations among the OAM components and among the angular positions of pseudothermal light. Furthermore, we show that the presence of these correlations leads to a variety of complex interference structures that correspond to the azimuthal analog of the HBT effect. In the original HBT experiment, two detectors were used at different locations to gain information about the physical size of a distant incoherent source. In our experiment, we use two detectors to measure intensity correlations between two OAM components of an incoherent source with controllable spatial and temporal coherence. We show that such correlations unveil the azimuthal structure of the source, which is shaped in the form of double angular slits in our realization. We study the far-field pattern by projecting it onto various OAM modes, and measure first- and second-order interference patterns of this structure. We identify two key signatures of the azimuthal HBT effect. The first is that HBT interference can show features in the OAM mode distribution at both the frequency and at twice the frequency of the first-order coherence produced by coherent light. The second consists of a shift of the interference structure when plotted as a function of OAM.  We find that each of these effects depends on the strength of the fluctuations of the pseudothermal light. We also study the nature of the correlations between different OAM components and between different angular positions of pseudothermal light, and we find that these depend on the strength of the fluctuations as well. These effects correspond to the classical counterpart of  azimuthal Einstein-Podolsky-Rosen (EPR) correlations \cite{Leach:2010bm}, and throughout this article, we highlight the similarities and differences between thermal and quantum correlations as manifested in the azimuthal degree of freedom.

\section{Results}
\section{Origin of HBT interference in the OAM domain}

As in the original HBT experiment, we collect light from two portions of a random field. This is carried out through the use of two angular slits. We represent the optical field after the slits as 
\begin{equation}
\Psi(r,\phi) = \mathcal{E}(r)\Phi(r,\phi)[A(\phi)+A(\phi-\phi_0)].
\end{equation}
Here,  $\mathcal{E}(r)$ represents the coherent optical field produced by a  
laser, $\Phi(r,\phi)$ is a particular realization of a random phase screen, and $A(\phi)$ describes the transmission function of the angular slits. $A(\phi)$ is centered at 0 radians, and, therefore, $A(\phi-\phi_0)$ is centered at $\phi_0$. We next consider the projection of the optical field of Eq.\ (1) onto a set of OAM modes. The result of such a measurement is described by the quantity $a_{p\ell}$ defined as $\int r \, dr \, d\phi \, (2\pi)^{-1/2} R_{p}^{*}(r)e^{-i\ell\phi}\Psi(r,\phi)$, where $R_{p}^{*}(r)$ is a radial mode function with  radial index $p$ and $\ell$ is the OAM index.  Consequently, the measured intensity for each OAM projection $I_\ell$ is equal to $\sum_p\abs{a_{p\ell}}^2$.  The average of the intensity over an ensemble of different realizations of the fluctuating field is then given by 
\begin{equation} \label{eq1}
\langle I_{\ell}\rangle  = \frac{\alpha^2\text{sinc}^2 \left(\alpha\ell/2\right)}{2\pi^2} \int r \, dr \abs{\mathcal{E}(r)}^{2}\{ 2+e^{-i\ell\phi_{0}} \langle \Phi^{*}(r,0)\Phi(r,\phi_{0})\rangle+ e^{i\ell\phi_{0}} \langle \Phi^{*}(r,\phi_{0}) \Phi(r,0) \rangle\},
\end{equation}
where $\alpha$ is the width of the slits, and the ensemble average is denoted by $\langle...\rangle$. It is evident that the angular double slit gives rise to Young's (first-order) interference in the OAM-mode distribution of the optical field and that this interference is dependent on the angular separation of the two slits, $\phi_0$.  Furthermore, the visibility of the interference pattern is determined by the terms $\langle \Phi^{*}(r,0)\Phi(r,\phi_{0})\rangle$ and $\langle \Phi^{*}(r,\phi_{0})\Phi(r,0)\rangle$, which quantify the field correlation between two different angular positions. These terms are sensitive to the phase difference of the field at two points. Consequently, the interference visibility becomes smaller as the degree as spatial coherence is reduced.

In direct analogy to the HBT experiment, in which two detectors measure the transverse momentum (far-field) distribution of a random field emitted from two locations of a star, we measure the correlation between two OAM components of light emitted from a random source shaped as two angular slits. Similar to linear position and linear momentum, angular position and OAM are conjugate variables and form a Fourier pair. Thus, we consider the second-order coherence function $G^{(2)}_{\ell_1,\ell_2} = \langle I_{\ell_1}I_{\ell_2} \rangle$, which is the key quantity that describes the azimuthal HBT effect. This quantity is a measure of the intensity correlations between the components of the the field with OAM values $\ell_1$ and $\ell_2$.  

We consider a special case in which we measure the the second-order correlation at symmetrically displaced OAM values of $\ell$ and $-\ell$.  In the context of the original experiment of HBT, this situation would involve measuring the receiving apertures by equal amounts in opposite directions.   To analyze this situation, we need to determine the second-order coherence function $G^{(2)}_{\ell,-\ell} = \langle I_{\ell}I_{-\ell} \rangle$.  We find that this quantity can be expressed (see Supplementary Materials) as
\begin{align}
\langle I_{\ell}I_{-\ell} \rangle&= \mathcal{G}_{0}+\mathcal{G}_{\ell}+\mathcal{G}_{2\ell}.
\end{align}
The intensity correlation function thus consists of three contributions. The first is a constant term denoted by $\mathcal{G}_{0}$ whose form is shown in the Supplementary Materials. The second term, $\mathcal{G}_{\ell}$, describes an interference pattern that oscillates in $\ell$ at the same frequency as $\langle I_\ell \rangle$ and is given by
\begin{align}
\mathcal{G}_{\ell} = \frac{\alpha^2\sinc^2(\alpha\ell/2)}{2\pi^2} \int r_1 dr_1r_2 dr_2 \abs{\mathcal{E}(r_1)}^{2}\abs{\mathcal{E}(r_2)}^{2}\big(e^{-i\ell\phi_{0}}\{ \langle \Phi^{*}(r_1,0)\Phi(r_1,\phi_{0})\rangle + \langle \Phi^{*}(r_2,\phi_{0}) \Phi(r_2,0) \rangle\} + \rm{c.c.}\big).
\end{align}
The last term, $\mathcal{G}_{2\ell}$, shows an interference pattern that oscillates in the OAM value $\ell$ with twice the frequency of $\langle I_\ell \rangle$ and it is given by
\begin{align} \label{eq2}
\mathcal{G}_{2\ell} & = \frac{\alpha^4\text{sinc}^4 \left(\alpha\ell/2\right)}{4\pi^4}\int r_1 dr_1 r_2 dr_2 \abs{\mathcal{E}(r_1)}^{2}\abs{\mathcal{E}(r_2)}^{2}\big(e^{-2i\ell\phi_{0}}\{ \langle \Phi^{*}(r_1,0)\Phi(r_1,\phi_{0})\Phi^{*}(r_2,\phi_{0}) \Phi(r_2,0) \rangle\} +\rm{c.c.}\big).
\end{align}

\noindent  We see that the contribution  $\mathcal{G}_{\ell}$ depends on a phase-sensitive term $\langle \Phi^{*}(r,0)\Phi(r,\phi_{0})\rangle$ that decreases in magnitude with increasing randomness induced by field fluctuations.  The visibility of this contribution to the interference pattern thus decreases with increasing field fluctuations.  However the contribution $\mathcal{G}_{2\ell}$ is proportional to a positive-definite quantity $\langle \abs{\Phi(r,0)}^2\abs{\Phi(r,\phi_{0})}^2 \rangle$ that survives even in the presence of the fluctuations in the chaotic field.

\begin{figure*}[h]
    \centering
    \includegraphics[width=0.75\textwidth]{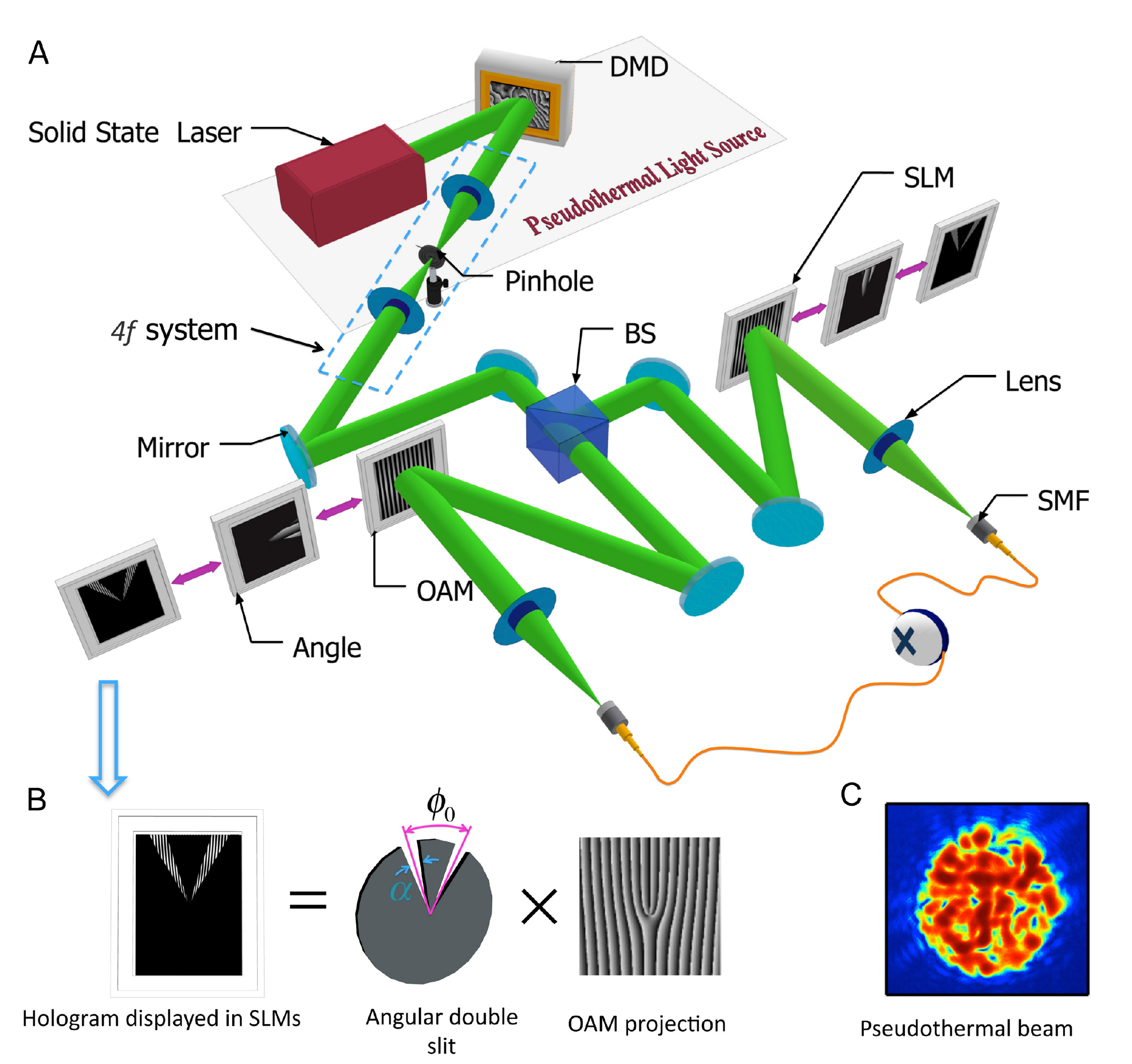}
    \caption{Experimental setup for the study of the azimuthal HBT effect.  (A) The 532 nm output of a solid laser is directed onto a digital micro-mirror device (DMD), where a random transverse phase structure is impressed onto the beam. A 4$f$ optical system consisting of two lenses with different focal lengths (figure not to scale) and a pinhole is used to isolate the first diffraction order from the DMD, which is a pseudothermal beam of light. This beam is then passed through a beam splitter (BS) to create two identical copies.  Each copy is sent to a separate spatial light modulator (SLM) onto which a computer-generated hologram is encoded.  (B) For the HBT measurements, a pair of angular slits is encoded onto the SLMs.  In addition, forked holograms corresponding to OAM values are encoded onto the same holograms to project out controllable OAM components.  For our measurements of the OAM and angular-position correlation functions, we do not use the double slit but simply project onto OAM values or angular wedges, respectively. (C) Intensity distribution of a generated pseudothermal beam of light. }
    \label{fig:setup}
\end{figure*} 

\section{Experimental demonstration of azimuthal HBT interference}

Our experimental setup is depicted in Fig.\ 1 (A and B). We use a solid state laser working at 532 nm along with a digital micro-mirror device (DMD) and a 4$f$ optical system containing two lenses and a spatial filter in the Fourier plane to isolate one order of diffraction from the DMD.  We first impress a sequence (at a 1.4-kHz writing rate) of random transverse structures having Kolmogorov statistics onto the beam to simulate thermal light  \cite{Harding:1999uv, Mirhosseini:2013go}. For details, see Materials and Methods. This procedure modifies the spatial and temporal coherence of the beam in a  fashion similar to the modification induced by a rotating ground glass plate \cite{Arecchi:1965vp} (see the intensity distribution of the beam in Fig. 1 C), which is often used to produce light with thermal statistics.  We quantify the spatial coherence of the beam by means of the Fried coherence length $r_0$\cite{Glindemann:1993jb}.  The strength of spatial phase variations within the beam increases as $r_0$ decreases. By virtue of ergodicity, iterating through an ensemble of such holograms results in random phase fluctuations in time characterized by the parameter $r_0$. The structured beam is then split into two parts at a beam splitter, and each is imaged onto a spatial light modulators (SLM).  On each SLM, a pair of angular slits and a forked diffraction grating are encoded (see Fig.\ 1B).  The first diffraction order of the SLM is collected by a single-mode optical fiber (SMF), measured by avalanche photodiodes (APDs), and their degree of correlation is then computed. The time window for determining coincidence events is set to 42 ns, and the total accumulation time is set to 15 s. 

 We begin with the measurement of first-order (Young's) interference in the OAM domain, which can be observed in the OAM-mode distribution of light measured by either of the two detectors. For each value of $\ell$, we impress several hundred random phase screens onto the DMD, all  characterized by the same value of $r_0$, and we then calculate the correlation of the intensity.  We  repeat the experiment for all $\ell$ in the range $\ell=-15$ to $\ell=+15$. We perform this task by encoding holograms onto the SLMs in which the two angular apertures are multiplied by different forked diffraction gratings (see Fig. 1B). The OAM-mode distributions of the field as given by $\langle I_\ell \rangle$ are shown in Fig.\ 2 (A to D).  Fig.\ 2A shows the interference obtained when spatially coherent light is used, and Fig.\ 2 (B to D) shows the interference for different regimes of pseudothermal light, as characterized by successively decreasing values of $r_0$. The visibility is seen to decrease with the decrease of the spatial coherence of the source.

\begin{figure*}[ht]
    \centering 
    \includegraphics[width=1.01\textwidth]{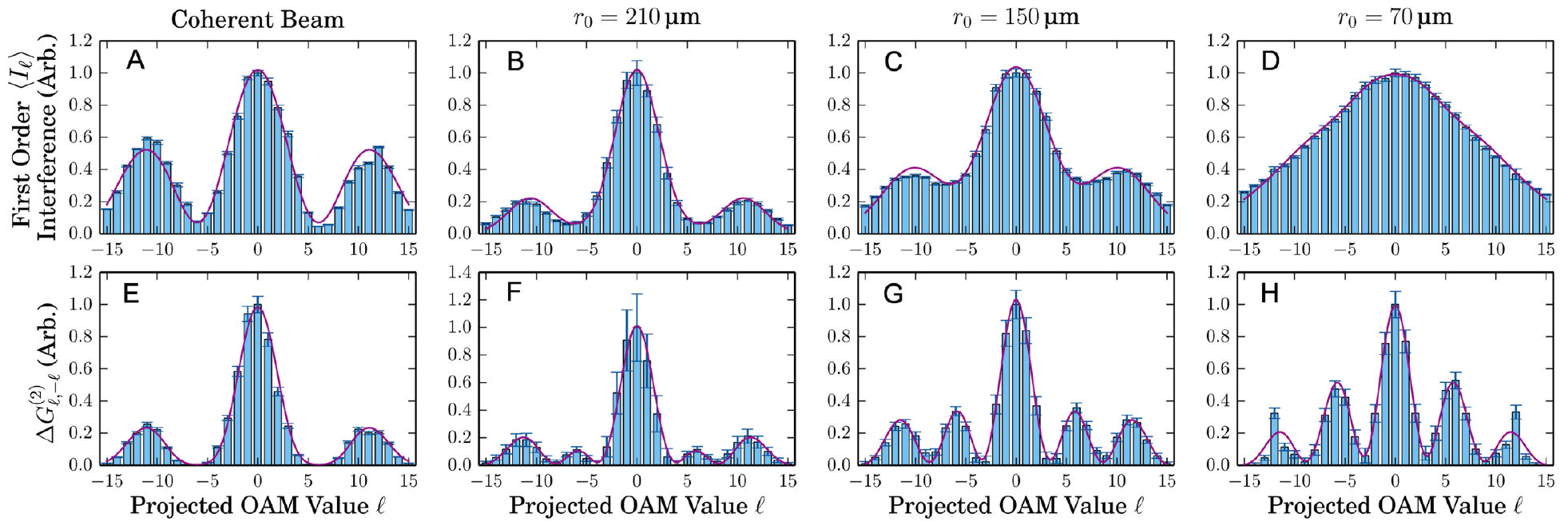}
    \caption{
    Interference transitions in the OAM-mode distribution of light. (A to D) First-order (Young's) interference. (E to H) Second-oder HBT interference. The first column (A and E) shows interference produced by coherent light, whereas the other panels show the measured interference for different strengths of the fluctuations of pseudothermal light, as characterized by the Fried coherence length. In each case, the angular width of the slits $\alpha$ is $\pi/12$ and the angular separation of the slits $\phi_0$ is $\pi/6$. Bars represent data, whereas the line is the theoretical curve predicted by theory.}
    \label{fig:setup}
\end{figure*} 

We next study second-order coherence. Our  experimental results for the second-order coherence function $\Delta G^{(2)}_{\ell,-\ell}$, defined as $\mathcal{G}_{\ell} + \mathcal{G}_{2\ell}$,   are shown in Figs.\ 2 (E to H).  For a coherent beam (Fig. 2E), $\mathcal{G}_{\ell}$ is the dominant contribution to $\Delta G^{(2)}_{\ell,-\ell}$.  We reach this conclusion by noting that the data oscillate at the same frequency as the first-order interference shown in Fig. 2A and by recalling the discussions of Eqs.\ (4) and (5).   We also note that  $\mathcal{G}_{\ell}$ decreases as the degree of the spatial coherence of the source is reduced, making $\mathcal{G}_{2\ell}$ the dominant contribution in this case; we reach this conclusion by an examination of Eq.\ (5),  which shows that $\mathcal{G}_{2\ell}$, in contrast to $\mathcal{G}_{\ell}$, does not decrease with decreasing degree of spatial coherence of the source. We see this behavior in the sequence of results shown in Figs. 2 (F to H).  For example, in Fig. 2F, the contribution from $\mathcal{G}_{2\ell}$ is smaller than that from $\mathcal{G}_{\ell}$. This transition is marked by the formation of second-order correlations in the angular position and OAM variables.

 It is interesting that there is a regime of random fluctuations for which strong frequency-$\ell$ oscillations are seen in the first-order interference while strong frequency-$2 \ell$ oscillations are seen in the second-order interference (see Figs.\ 2, B and F).  Note also that, for the case of quantum correlations, entangled photons do not produce interference in singles but only in correlations such as those shown in Fig.\ 2 (D and H) \cite{Jha:2010gf, Xiong:2005kj}. The interplay between $\mathcal{G}_{\ell}$ and $\mathcal{G}_{2\ell}$ might be useful to the study of the relationship between coherence and the quantum nature of light. 

It is important to remark that different degrees of coherence define regimes of the HBT effect \cite{Hanbury:1958cu}, as shown in Fig. 2. In our case, the varying relative magnitude of the three terms contributing to the second-order coherence $\mathcal{G}^{(2)}_{\ell,-\ell}$ results in different shapes (see Eq. 3). For example, $\mathcal{G}_{2\ell}$ makes the pattern in Fig. 2E sharper, but the same term changes the frequency of the interference structure in Fig. 2H.

The general form  of the azimuthal HBT effect is obtained when the intensity correlations are calculated for arbitrary mode indices $\ell_1$ and $\ell_2$. As discussed above, the HBT effect depends on the degree of coherence of the source. Specifically, an interesting feature is observed for the partially coherent regime characterized by $r_0$ equal to 150 $\mu$m. In our experimental study of this situation, we hold the OAM value measured in one arm of our interferometer fixed at the value $\ell_0$ whereas we vary the OAM value in the other arm. We set the value of  $\ell_0$ first to $+2$ and later to $-2$.  In the other arm, we perform measurements for each value in the range $\ell = -15$ to $+15$.  The results of these measurements are shown in Fig.\ 3.  It should be noted that the OAM spectrum plotted as a function of the the OAM value of arm 2 is shifted left (see Fig. 3A) or right (Fig. 3B) depending on the value of OAM chosen for arm 1.  The procedure used in the measurement is analogous to using one fixed detector and one moving detector in the  original setup of HBT \cite{HBT1}. The results of Fig.\ 3 (A and B) are described by the quantity $\langle I_\ell I_{\ell_0}\rangle$ and can be expressed in terms of five contributing terms (see section 2 of the Supplementary Materials). 

For the strength of fluctuations that we used for these measurements, one of the detectors measures an interference pattern equal to the one shown in Fig.\ 2C, whereas the other measures a noisy but constant signal.  When the correlation of the two  signals is calculated, the visibility of the interference pattern is dramatically increased and shifted in the OAM-mode distribution of the field.  Effectively, we are using the random fluctuations of the field to increase the fringe visibility. For example, if instead of projecting an OAM value equal to 2 or -2 as we did, we  could project on $\ell$ equal to zero and retrieve the original but improved pattern with increased visibility. This effect could find importance in realistic applications. These effects manifest the presence of second-order correlations in the OAM components and angular positions of pseudothermal light.

\begin{figure}[h]
    \centering
    \includegraphics[width=1\textwidth]{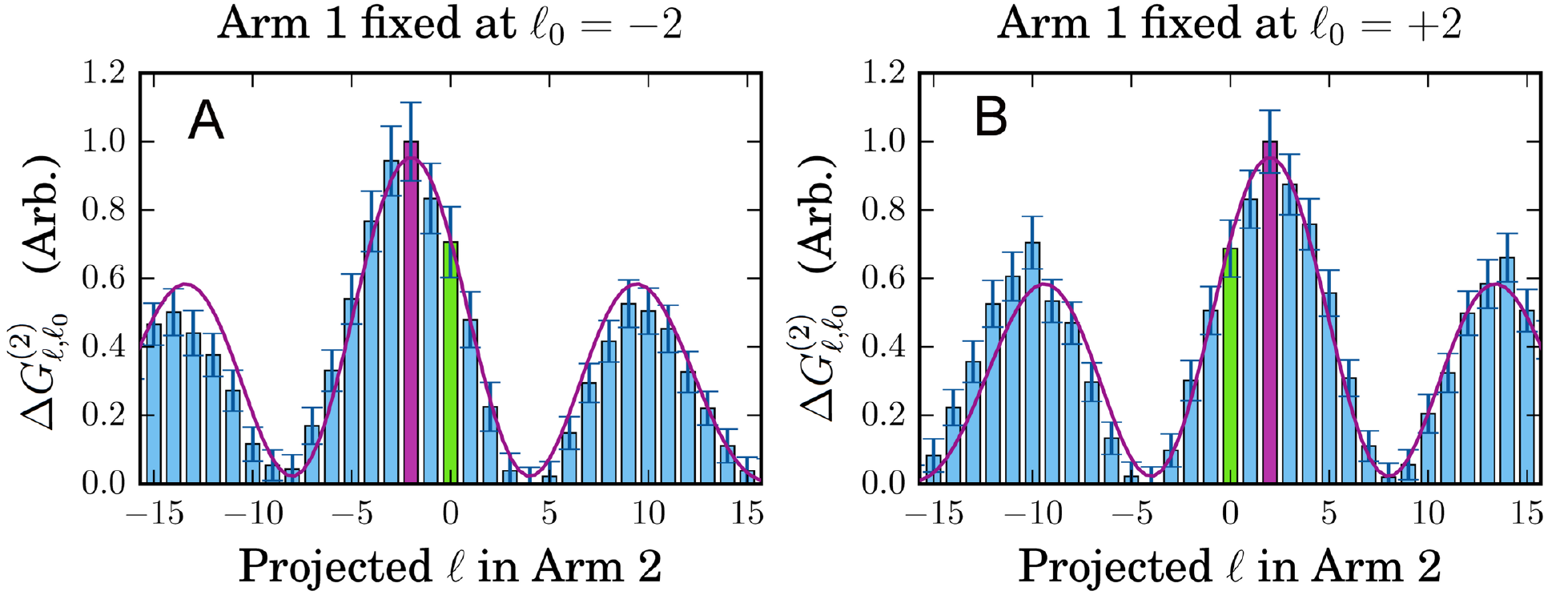}
    \caption{Experimental demonstration of the azimuthal HBT effect of light. (A and B) $\Delta G_{\ell,\ell_0}^{(2)}$ plotted as a function of the OAM value of arm 2 for two different values of the OAM number of arm 1. The green bar shows the center of the interference pattern for singles counts  shown in Fig. 2C, whereas the purple bar shows the center of the displayed interference pattern.}
    \label{fig:setup}
\end{figure}  

We would like to emphasize that although the angular slits and the forked holograms for OAM projections are realized on the same SLMs, they correspond to conceptually distinct components of the experiment. The angular slits are used to provide a nontrivial azimuthal structure for the incoherent source, whereas the forked holograms are used to measure correlations in the OAM domain.

\section{Measurement of angular momentum correlations and angular position correlations}

Now we explore the nature of the underlying fluctuation-induced correlations in OAM and in angular position that lie at the origin of the HBT effect. The superposition of randomly fluctuating waves produces an OAM spectrum that broadens with the degree of fluctuation in the source of pseudothermal light. In the present experiment, the OAM spectrum is controlled by setting $r_0$ equal to 70  $\mu$m. This situation produces a broad OAM spectrum that remains almost constant over the range of OAM values that we measure.  We use the same setup as that of Fig.\ 1, although we omit the two angular slits that we used in the studies of azimuthal HBT interference effects reported above.  On the first SLM (see Fig.\ 1), we display a forked hologram corresponding to a fixed value of OAM, whereas on the second SLM, we display a series of holograms with different values of OAM. The measured intensity for a single value of OAM $\langle I_\ell \rangle$ that is projected out using the SLM can be approximated as $\int r^2 dr^2 d\phi^2 \abs{\mathcal{E}(r)}^2g (r)^2$, where $g(r)$ is the Gaussian mode supported by the SMF (see section 3 of the Supplementary Materials).  

In Fig.\ 4A, we plot the measured value of $g^{(2)} = \langle I_{\ell_1}I_{\ell_2}\rangle/\langle I_{\ell_1}\rangle\langle I_{\ell_2}\rangle$.  We find a strong positive correlation between the OAM values measured in the two arms.   As shown in section 3 of the Supplementary Materials, in the limit of a strong fluctuations, second-order correlations in the OAM degree of freedom can be described by
\begin{equation} \label{eq4}
\langle I _{\ell_1} I_{\ell_2} \rangle  =\langle I_{\ell_1} \rangle \langle I_{\ell_2} \rangle(1+\delta_{\ell_1,\ell_2}).
\end{equation}
\noindent
Our experimental results show crosstalk between different OAM numbers that is not predicted by Eq.\ (\ref{eq4}).  This crosstalk results from experimental imperfections in the projective measurement process used to characterize OAM. The correlations in Fig. 4A show two significant differences from the quantum correlations observed in spontaneous parametric down conversion (SPDC). The first is that SPDC shows strong anti-correlations of the the two OAM values. This behavior is a consequence of the conservation of OAM in a parametric nonlinear optical process. The second difference is the presence of a background term (the ``1" in Eq.\ (\ref{eq4})), which prevents the existence of perfect correlations.

\begin{figure}[h]
    \centering
    \includegraphics[width=1\textwidth]{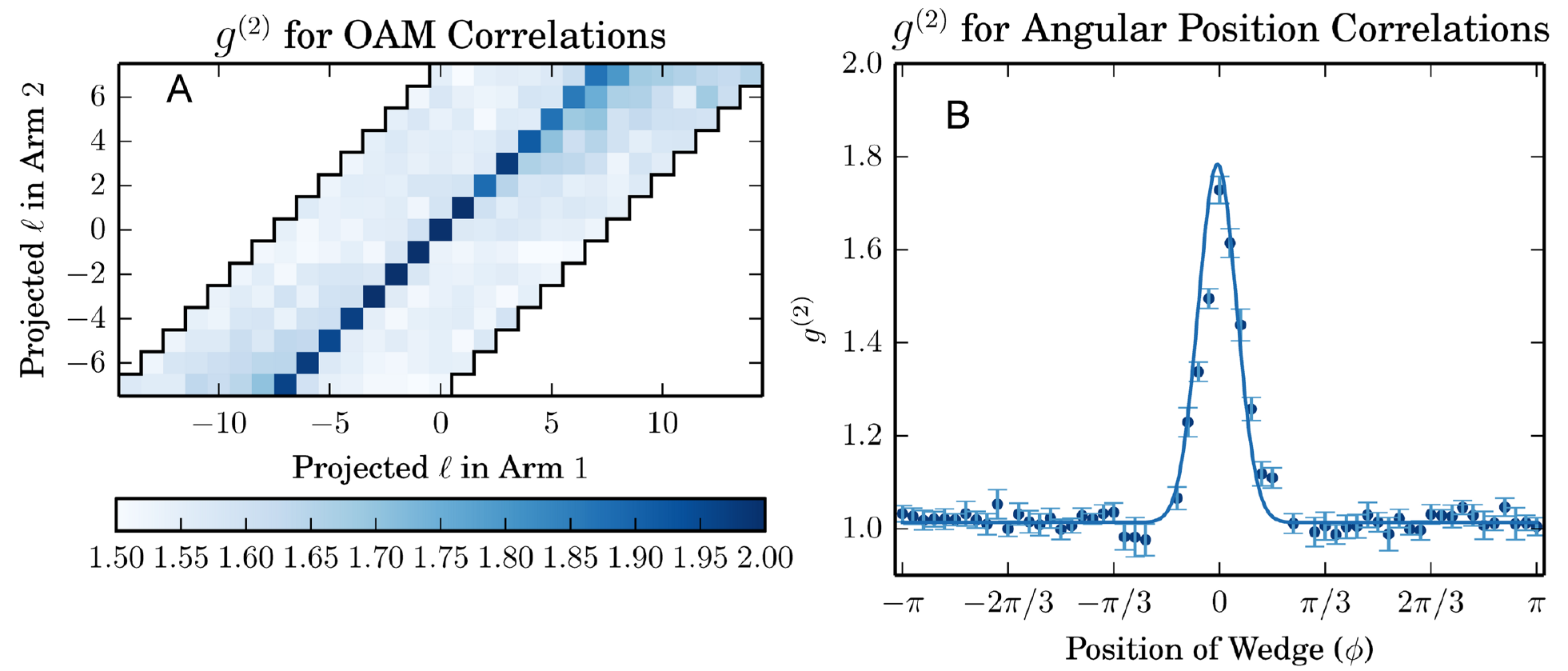}
    \caption{Measurement of intensity correlations in the angular domain for random light. (A) Normalized second-order correlation function in the OAM domain. (B) Presence of strong correlations for the conjugate space described by the angular position variable.}
    \label{fig:setup}
\end{figure}  

Randomly fluctuating beams also produce correlations in angular position. These correlations are investigated by encoding angular apertures onto the SLMs. In order to make our measurements precise, we utilize narrow angular apertures of $\pi/15$ radian size. We keep one aperture at a fixed location, and we measure correlations for $60$ different angular positions of the other aperture. Because of rotational symmetry, this procedure permits the full characterization of correlations in angular position. 
As shown in Fig. 4B, for this level of fluctuation, the intensities of the projected angular apertures are strongly correlated, and the nature of these correlations can be approximated by $\langle I_{\phi}I_{\phi_0}\rangle=\langle I_{\phi}\rangle \langle I_{\phi_0}\rangle(1+f(\phi-\phi_0))$. In this equation, the subscript $\phi$ refers to the arm where the variable-position angular aperture is placed and $\phi_0$ represents the arm with the fixed-position aperture.  Also, $f(\phi-\phi_0)$ represents some strongly peaked function (see the Supplementary Meaterials).  

As we have shown throughout this paper, the HBT correlations of pseudothermal light lead to effects that show resemblance to those previously observed with entangled photons \cite{Jha:2010gf, Leach:2010bm, Jack:2009wk, UribePatarroyo:2013fj}.  The reason for this behavior is that, in contrast to the degree of second-order coherence that describes coherent light, the functions that describe second-order correlations in angular position and OAM for random fields are nonseparable. For example, Eq.\ 6 does not contain the product of the averaged intensities measured by each of the two detectors. The presence of a term that describes point-to-point correlations (in this case, the delta function $\delta_{\ell_1,\ell_2}$) does not allow the factorization of the degree of coherence as the simple product of intensities between the two arms. As a consequence, the HBT structures are also described by a nonseparable function, and its frequency, visibility, and shifts increase with the fluctuations of the source or the strength of angular position and OAM correlations. As the strength of the fluctuations decreases, the nonseparable part of the function tends to vanish, and thus, the second order correlation function can be factorized in terms of OAM or angular position. A separable function will not lead to the HBT effect in the OAM-mode distribution of light; see the transition shown in Fig. 2.

Intensity correlation in the OAM components and angular position of pseudothermal light show similarities with the azimuthal EPR effect, observed in photons entangled in angular position and OAM \cite{Leach:2010bm}. However, our results show that for pseudothermal light, the correlations are present but not perfect, unlike the case of entangled photons where the correlations are perfect. Thus, it is impossible to violate, for example, the azimuthal EPR criterion $(\Delta\ell)^2(\Delta\phi)^2\geq1/4$. However, as shown in Fig. 4, our correlations are stronger for same values of OAM or angular positions. For example, if background subtraction is performed, the variance product for $\Delta\ell$ and $\Delta\phi$ is similar to that achieved for nonclassical light. For our experimental results $(\Delta\ell)^2(\Delta\phi)^2$ is $0.054$, of similar order to the one reported by Leach $et$ $al.$ \cite{Leach:2010bm}. The uncertainties were measured by performing a least squares fit of the data to a Gaussian distribution and recording the standard deviation of the result. Note that this does not imply a violation of the EPR criterion. 

\section{Discussion}
The azimuthal HBT effect unveils  fundamental physics that can be applied to develop novel applications that exploit OAM correlations in random light. We believe that many interesting protocols for remote sensing and object identification that use azimuthal correlations in entangled photons will be able to exploit azimuthal correlations in random light and the azimuthal HBT effect \cite{Jack:2009wk, UribePatarroyo:2013fj,  Chen:2014gf}. Furthermore, in recent years, researchers have developed interest in utilizing beams carrying OAM for applications in astronomy, but unfortunately the propagation through random media produces chaotic phase fluctuations and optical vortices \cite{Chen:2014gf, Paterson:2005vt, Gbur:2003wm, Kumar:2012wk, Dennis:2010ez}. These effects pose serious problems for methods based on OAM of light, limiting their applications \cite{Hetharia:2014wl, Yao:2011uo}. However, it has been shown that second-order interference effects are less sensitive to the coherence properties of the source. This is one of the advantages of the HBT interferometer against the Michelson interferometer \cite{Wolf:1999cu}. In addition, it has been demonstrated that imaging schemes based on second-order correlations are robust against turbulence \cite{Dixon:2011tn}. Therefore, we suggest that the azimuthal HBT effect offers the possibility of exploring novel phenomena in astrophysics, one example being the relativistic dynamics produced by rotating black holes \cite{Tamburini:2011wm}.

 We have demonstrated that random fluctuations of light give rise to the formation of intensity correlations in the OAM components and angular positions of pseudothermal light. These correlations are manifested through a new family of interference structures in the OAM-mode distribution of pseudothermal light that can be described by the azimuthal HBT effect. We have shown how the strength of the random fluctuations of light determines various regimes for this effect. In addition, we identified two key features of the azimuthal HBT effect. The first is characterized by a structure in which the OAM frequency is doubled with respect to the interference produced by a coherent beam of light. The second is marked by a shift of the OAM spectrum with a change in the OAM reference value. We anticipate that these properties of random optical fields will be fundamentally important for applications where quantum entanglement is not required and where correlations in angular position and OAM suffice. 
 \newpage
\vspace{4mm} \section{Materials and Methods}
\vspace{4mm}

 \textbf{Source of Pseudothermal light. }  Pseudothermal light was generated by means of phase screen holograms obeying Kolmogorov statistics. Kolmogorov's statistical theory is used to model chaotic turbulent fluids. We have generated Kolmogorov phase screens for varying levels of simulated randomness by using the approximate power spectral density of $\Phi(f) \approx 0.023 r_0^{-5/3} f^{-11/3}$. Fried's parameter $r_0$ is related to the average coherence length between two points in the beam. By adjusting Fried's parameter $r_0$, we can increase or decrease the size of and the distance between the phase cells and thus the amount of randomness in the phase screens. By adding normally distributed deviations to the power spectral density, we can then take the real part of the inverse Fourier transform in order to generate a single Kolmogorov phase screen. % http://www.activeopticalsystems.com/docs/AN021_Kolmogorov%20Spectrum%20Turbulence%20Analysis.pdf

A DMD can be used to manipulate both the phase and amplitude profile of a light beam. A translation in a binary diffraction grating will cause a phase shift to occur in the diffracted light, whereas varying the duty cycle of the periodic grating will change the efficiency, and thus the amplitude, of the diffracted beam. Both of these techniques can be done locally to spatially control the phase and amplitude of the beam. The generated Kolmogorov screens were then converted into binary diffraction gratings to be displayed on a DMD.

\begin{figure}[h]
  \centering
  \setlength{\fboxrule}{0.7pt}
  \fbox{\includegraphics[width=0.19\columnwidth]{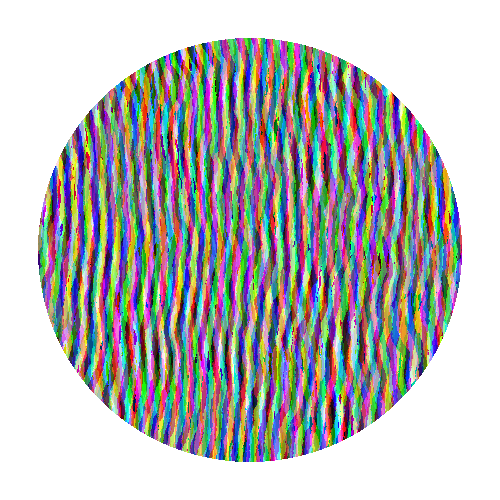}}
  \caption{Example of a frame sent to the DMD.  It contains 24 binary holograms encoded in bit plane slices.}
  \label{fig:composite}
\end{figure}

We used the Texas Instruments LightCrafter Evaluation Module (DLPC300) which drives a Texas Instruments DLP3000 DMD. The DMD contains an array of $608\times684$ micromirrors with a total diagonal length of \SI{7.62}{\milli\meter}. The DMD was operated in a mode that allowed a binary pattern to be displayed at a rate of \SI{1440}{\hertz}. The DMD takes a 24-bit color \SI{60}{\hertz} signal over an HDMI connection. Because the image contains 24 bits, a single video frame can contain 24 binary images. In this mode, the DMD will cycle through the least significant bit to the most significant bit in the blue signal of a frame. Then, the DMD will display the bits in the red signal and, finally, the green signal. Kolmogorov screens (72,000) were encoded into three thousand 24-bit frames for each value of Fried's parameter $r_0 = $ 70 $\mu$m, 150 $\mu$m and 210 $\mu$m. Figure 5 shows an example of one of the generated frames sent to the DMD. This frame contains 24 binary holograms encoded in the bit planes of the image to be displayed sequentially.

\begin{figure}[h]
  \centering
  \includegraphics[width=0.5\columnwidth]{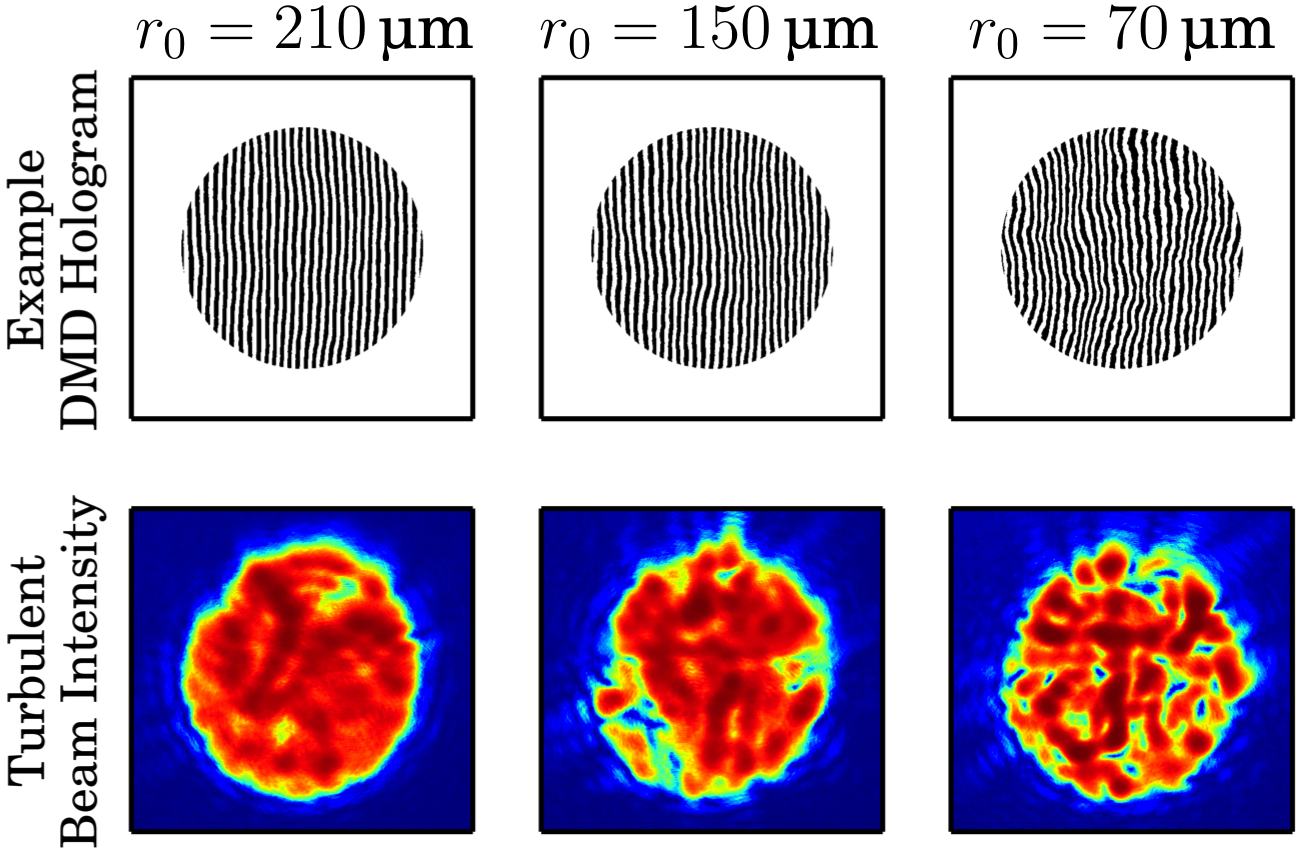}
  \caption{Example DMD holograms and resulting beams measured before the image plane.}
  \label{fig:beam}
\end{figure}

Figure 6 shows examples of the intensity distribution for three random beams generated by this method. In addition, in each case, an example of one of the holograms used to produce the beam is shown.
Note that the randomness within the beam increases as the value of $r_0$ decreases. 

 \newpage
\vspace{4mm} \section{Supplementary Materials}
\vspace{4mm}

\textbf{1. The HBT effect for symmetrically displaced modes ($\ell$ and $-\ell$).}

Here, we derive the equations utilized in the manuscript. We start by describing HBT interference of pseudothermal light.  We assume that the beam of light from our laser is described by the electric field $\mathcal{E}(r)$, where $r$ is the radial coordinate in the transverse plane. In addition, we assume that the initial electric field does not possess any azimuthal dependence. We encode a random Kolmogorov phase screen $\Phi(r,\phi)$ onto the beam. Later, the field illuminates two angular apertures centered on  angles $0$ and $\phi_{0}$. Thus the field after the two slits is given by 
\begin{equation}
\begin{split}
\Psi(r,\phi)=\mathcal{E}(r)\Phi(r,\phi)[A(\phi)+A(\phi-\phi_{0})].
\end{split}\label{eq:1}.
\end{equation}
As described earlier in the manuscript, we replace the widely used ground-glass plate with a series of phase screens that change rapidly in comparison to the accumulation time of the measurement, thus creating an ensemble of field realizations. The next step is to find the intensity of the field for a given OAM eigenstate. We can write the electric field after the two slits as a linear combination of radial-OAM modes. We designate a complete radial basis as $R_{p}(r)$, although we do not make use of any explicit form for this basis. We thereby express the field after the slits as
\begin{equation}
\Psi(r,\phi)=\sum_{\ell,p} a_{p\ell}R_{p}(r)\frac{e^{i\ell\phi}}{\sqrt{2\pi}},
\end{equation}
where the expansion coefficients $a_{p\ell}$ are given by
\begin{equation}
a_{p\ell}=\int r dr d\phi R_{p}^{*}(r)\frac{e^{-i\ell\phi}}{\sqrt{2\pi}}\Psi(r,\phi).
\end{equation}
Thus, the measured intensity after projecting the beam onto  OAM mode $\ell$ is given by 
\begin{equation}
\begin{split}
I_{\ell}&=\sum_{p} |a_{p\ell}|^{2}=\sum_{p} \int r_{1} dr_{1} r_{2} dr_{2} d\phi_{1} d\phi_{2} \Psi^{*}(r_{1},\phi_{1})R_{p}(r_{1})\frac{e^{i \ell \phi_{1}}}{\sqrt{2\pi}}\frac{e^{-i\ell\phi_{2}}}{\sqrt{2\pi}}R_{p}^{*}(r_{2})\Psi(r_{2},\phi_{2}) \\
&=\int r_{1} dr_{1} r_{2} dr_{2} d\phi_{1} d\phi_{2} \Psi^{*}(r_{1},\phi_{1}) \Psi(r_{2},\phi_{2})\frac{e^{i\ell(\phi_{1}-\phi_{2})}}{2\pi}\sum_{p}
R_{p}(r_{1})R_{p}^{*}(r_{2})\\
&=\frac{1}{2\pi}\int r dr d\phi_{1} d\phi_{2} \Psi^{*}(r,\phi_{1}) \Psi(r,\phi_{2})e^{i\ell(\phi_{1}-\phi_{2})},
\end{split}
\end{equation}
where the last form comes from using the relation $\sum_{p} R_{p}(r_{1})R_{p}^{*}(r_{2})= (1/r_1)\delta(r_{1}-r_{2})$, which is true for any complete normalized set of basis functions,  where $\delta (r)$ is the usual Dirac delta function. Now we replace $\Psi(r,\phi)$ with the electric field after the two angular slits given by Eq.\ \ref{eq:1}. For simplicity, we first approximate $A(\phi)$ by $\delta(\phi)$ and $A(\phi-\phi_0)$ by $\delta(\phi-\phi_0)$. The  quantity $I_\ell$ then becomes 
\begin{equation}
\begin{split}
%I_{\ell}=\frac{1}{2\pi}\int d\phi_{1} d\phi_{2} r& dr e^{{i\ell(\phi_{1}-\phi_{2})}}
%\abs{\mathcal{E}(r)}^{2} \Phi(r,\phi_{1})\Phi^{*}(r,\phi_{2})\\
%&\left\{ \delta(\phi_{1})\delta(\phi_{2})+\delta(\phi_{1})\delta(\phi_{2}-\phi_{0})+
 %\delta(\phi_{1}-\phi_{0})\delta(\phi_{2})+\delta(\phi_{1}-\phi_{0})\delta(\phi_{2}-\phi_{0}) \right\}\\
 I_{\ell}=\frac{1}{2\pi}\int r dr \abs{\mathcal{E}(r)}^{2} & \left\{ 2+e^{-i\ell\phi_{0}} \Phi^{*}(r,0)\Phi(r,\phi_{0}) +e^{i\ell\phi_{0}} \Phi^{*}(r,\phi_{0}) \Phi(r,0)  \right\}.
\end{split} 
\end{equation}
We next take the ensemble average to obtain
\begin{equation}
\begin{split}
\langle I_{\ell}\rangle=\frac{1}{2\pi}\int r dr \abs{\mathcal{E}(r)}^{2} & \left\{ 2+e^{-i\ell\phi_{0}} \langle \Phi^{*}(r,0)\Phi(r,\phi_{0})\rangle  +e^{i\ell\phi_{0}} \langle \Phi^{*}(r,\phi_{0}) \Phi(r,0) \rangle \right\}.
\end{split}\label{eq:6}
\end{equation}
In reality, however, the finite size of the slits produces an envelope, caused by diffraction, that modulates the form of the interference pattern. If a slit with a width $\alpha$ is considered, the interference pattern can be easily calculated to be $\int_{-\pi}^{\pi} d\phi f(\phi)e^{-i\phi\ell}$, where in our case $f(\phi)$ is equal to $1$ in the range from $-\alpha/2$ to $\alpha/2$ and is equal to 0 otherwise. This integral produces a diffraction envelope given by $\frac{\alpha}{2\pi}\text{sinc} \big( \frac{\ell\alpha}{2} \big)$. The diffraction produced by the second slit can likewise be described as $\frac{\alpha}{2\pi}\text{sinc} \big( \frac{\ell\alpha}{2} \big) e^{-i\ell\phi_0}$. The intensity of the total diffraction is described as $\big(\frac{\alpha}{\pi}\big)^2\text{sinc}^2 \big( \frac{\ell\alpha}{2} \big)(1+\cos{(\ell\phi_0)})$. Taking this result into account, we find that the first-order-interference diffraction pattern is given not by Eq.\ \ref{eq:6}. but rather by 
\begin{equation}
\langle I_{\ell}\rangle=\frac{\alpha^2\text{sinc}^2(\ell\alpha/2)}{2\pi^2}\int r dr \mathcal{E}^{2}(r)  \left\{ 2+e^{-i\ell\phi_{0}} \langle \Phi^{*}(r,0)\Phi(r,\phi_{0})\rangle+e^{i\ell\phi_{0}} \langle \Phi^{*}(r,\phi_{0}) \Phi(r,0) \rangle \right\}. 
\end{equation}

We next develop appropriate approximations for the quantities defined above. A reasonable assumption is that the field fluctuations follow Gaussian statistics such that 
\begin{equation}
\begin{split}
\langle \Phi^{*}(r_1,0)\Phi(r_2,\phi_{0})\rangle=\exp{(-\frac{r^2_1+r^2_2-2r_1r_2\cos\phi_0}{r^2_0})}. 
\end{split}\label{eq:8}
\end{equation}
By setting $r_1$ equal to $r_2$, we find that 
\begin{equation}
\begin{split}
\langle \Phi^{*}(r,0)\Phi(r,\phi_{0})\rangle=\exp({-\frac{4 r^{2}|\sin\frac{\phi_{0}}{2}|^{2}}{r_{0}^{2}}})=\exp(-\beta r^{2}).
\end{split}\label{eq:9}
\end{equation}
The last form of this expression defines the quantity $\beta$.  For a fully coherent beam, (that is, for $r_{0}\gg r$, where  $r_0$ is the Fried parameter introduced in Section 3) we see that to very high accuracy $\langle \Phi^{*}(r,0)\Phi(r,\phi_{0})\rangle$ is equal to 1. As $r_{0}$ decreases the value of the correlation function $\langle \Phi^{*}(r,0)\Phi(r,\phi_{0})\rangle$ also decreases.  Through use of Eq.\ \ref{eq:8}  expression (6) for the intensity can be expressed as 
\begin{equation}
\begin{split}
\langle I_{\ell}\rangle&=\mathcal{I}_{0}+\mathcal{I}_{\ell}\\
\mathcal{I}_{0}&=\frac{1}{\pi}\int r dr \mathcal{E}^{2}(r)\\
\mathcal{I}_{\ell}&=\frac{\cos(\ell\phi_0)}{\pi} \int r dr \abs{ \mathcal{E}(r)}^{2} \exp(-\beta r^{2}).
\end{split}
\end{equation}
As $r_{0}$ decreases, $\beta$ increases and most of the contribution to the integral in the last expression comes from $r\approx 0$, but since the integrand is zero at that point, the integral vanishes. In this limit, $\mathcal{I}_{0}$ makes the only contribution to  $\langle I_{\ell}\rangle$ and the spectrum becomes flat, that is, $\langle I_{\ell}\rangle$ shows no dependence on the value of $\ell$.  

Next we derive an expression for the correlations between projections onto OAM values of $\ell$ and $-\ell$, that is, 
\begin{equation}
\begin{split}
I_{\ell}I_{-\ell}&=\frac{1}{4\pi^{2}}
\int r_{1} dr_{1} \abs{\mathcal{E}(r_{1})}^{2}  \left\{ 2+e^{-i\ell\phi_{0}}  \Phi^{*}(r_{1},0)\Phi(r_{1},\phi_{0})  +e^{i\ell\phi_{0}}  \Phi^{*}(r_{1},\phi_{0}) \Phi(r_{1},0)  \right\} \\
&~~~~\times \int r_{2} dr_{2} \abs{\mathcal{E}(r_{2})}^{2}  \left\{ 2+e^{i\ell\phi_{0}}  \Phi^{*}(r_{2},0)\Phi(r_{2},\phi_{0})  +e^{-i\ell\phi_{0}}  \Phi^{*}(r_{2},\phi_{0}) \Phi(r_{2},0) \right\}\\
& \equiv \mathcal{G}_{0}+\mathcal{G}_{\ell}+\mathcal{G}_{2\ell},
\end{split} \label{eq:11}
\end{equation}
where the three contributions to $I_{\ell}I_{-\ell}$ are given by 
\begin{equation}
\begin{split}
\mathcal{G}_{0}=&\frac{1}{\pi^{2}}\int r_{1} dr_{1} r_{2} dr_{2} \abs{\mathcal{E}(r_{1})}^{2} \abs{\mathcal{E}(r_{2})}^{2} 
+\frac{1}{4\pi^{2}}\int r_{1} dr_{1} r_{2} dr_{2} \abs{\mathcal{E}(r_{1})}^{2} \abs{\mathcal{E}(r_{2})}^{2} \Phi^{*}(r_{1},0)\Phi(r_{1},\phi_{0}) \Phi^{*}(r_{2},0)\Phi(r_{2},\phi_{0}) \\
&+\frac{1}{4\pi^{2}}\int r_{1} dr_{1} r_{2} dr_{2} \abs{\mathcal{E}(r_{1})}^{2} \abs{\mathcal{E}(r_{2})}^{2}  \Phi^{*}(r_{1},\phi_{0}) \Phi(r_{1},0)  \Phi^{*}(r_{2},\phi_{0}) \Phi(r_{2},0)
\end{split} \label{eq:12}
\end{equation}
\begin{equation}
\begin{split}
\mathcal{G}_{\ell}=&\frac{1}{2\pi^{2}}\int r_{1} dr_{1} r_{2} dr_{2}\abs{\mathcal{E}(r_{1})}^{2} \abs{\mathcal{E}(r_{2})}^{2} e^{-i\ell\phi_{0}}  \left\{  \Phi^{*}(r_{1},0)\Phi(r_{1},\phi_{0}) + \Phi^{*}(r_{2},\phi_{0}) \Phi(r_{2},0)  \right\}+\text{c.c.},\text{ and}
\end{split}
\end{equation}
\begin{equation}
\begin{split}
\mathcal{G}_{2\ell}=&\frac{1}{4\pi^{2}}\int r_{1} dr_{1} r_{2} dr_{2}\abs{\mathcal{E}(r_{1})}^{2} \abs{\mathcal{E}(r_{2})}^{2} e^{-2i\ell\phi_{0}} \left\{ \Phi^{*}(r_{1},0)\Phi(r_{1},\phi_{0})\Phi^{*}(r_{2},\phi_{0}) \Phi(r_{2},0) \right\}+\text{c.c.}
\end{split}
\end{equation}
We next estimate the ensemble averages of these quantities. For a  field with strong random fluctuations,  the field correlation between two different angular positions is very small and  thus the term $\mathcal{G}_{\ell}$ does not contribute significantly to $\langle I_{\ell}I_{-\ell}\rangle$. A similar situation occurs for the second and third  contributions of $\mathcal{G}_{0}$;  
it is important to note that these terms 
contain the quantities $\langle\Phi^{*}(r_{1},0)\Phi(r_{1},\phi_0)\Phi^{*}(r_{2},0)\Phi(r_{2},\phi_0)\rangle$ and $\langle\Phi^{*}(r_{1},\phi_0)\Phi(r_{1},0)\Phi^{*}(r_{2},\phi_0)\Phi(r_{2},0)\rangle$, and these quantities vanish when $r_{1}=r_{2}$.  They vanish because they describe the average of the product of two chaotic and independent variables.  Thus, the main contributions to the second-order interference are the first term in Eq.\ \ref{eq:12} (which does not vary with $\phi_0$)  and the contribution 
\begin{equation}
\begin{split}
\langle\mathcal{G}_{2\ell} \rangle=&\frac{e^{-2i\ell\phi_{0}}}{4\pi^{2}}\int r_{1} dr_{1} r_{2} dr_{2}\abs{\mathcal{E}(r_{1})}^{2} \abs{\mathcal{E}(r_{2})}^{2}  \langle \Phi^{*}(r_{1},0)\Phi(r_{1},\phi_{0})\Phi^{*}(r_{2},\phi_{0}) \Phi(r_{2},0) \rangle+\text{c.c.}.
\end{split}
\end{equation}
\noindent
It is important to note  that,  contrary to the correlation functions given by $\langle\Phi^{*}(r_{1},0)\Phi(r_{1},\phi_0)\Phi^{*}(r_{2},0)\Phi(r_{2},\phi_0)\rangle$ and $\langle\Phi^{*}(r_{1},\phi_0)\Phi(r_{1},0)\Phi^{*}(r_{2},\phi_0)\Phi(r_{2},0)\rangle$, the  correlation function $\langle \Phi^{*}(r_{1},0)\Phi(r_{1},\phi_{0})\Phi^{*}(r_{2},\phi_{0}) \Phi(r_{2},0) \rangle$ is equal to unity for $r_1=r_2$.  We therefore obtain
\begin{equation}
\langle\mathcal{G}_{2l} \rangle=\frac{e^{-2i\ell\phi_{0}}}{4\pi^{2}}\int r_{1}^{2} dr_{1}\abs{\mathcal{E}(r_{1})}^{4}+\text{c.c.} \label{eq:16}
\end{equation}
We thus conclude that the quantity $\langle I_{l}I_{-l}\rangle$ is given by the sum of the contributions of Eq.\ \ref{eq:16} and the first term in Eq.\ \ref{eq:12}  or 
\begin{equation}
\langle I_{l}I_{-l}\rangle\approx  \left( \frac{1}{\pi}\int r dr \abs{\mathcal{E}(r)}^{2}\right)^{2} + \left( \frac{\cos2l\phi_{0}}{2\pi^{2}}\int r^{2} dr\abs{\mathcal{E}(r)}^{4} \right).
\end{equation}
For the case of slits with finite size, this result must be modified for the same reasons given in the discussion following Eq.\ \ref{eq:6}.  One thereby obtains 
\begin{equation}
\langle I_{l}I_{-l}\rangle\approx  \left( \frac{1}{\pi}\int r dr \abs{\mathcal{E}(r)}^{2}\right)^{2} + \frac{\alpha^4\text{sinc}^4 \left(\frac{\ell\alpha}{2}\right)}{4\pi^4} \left( \frac{\cos2l\phi_{0}}{2\pi^{2}}\int r^{2} dr\abs{\mathcal{E}(r)}^{4}\right).
\end{equation}

\textbf{2. The HBT effect for arbitrary mode indices $\ell_1$ and $\ell_2$.}

The intensity correlation between two arbitrary OAM modes $\langle I_{\ell_1}I_{\ell_2} \rangle$ produces a complicated second-order correlation function comprised of five terms. The contribution of each term is determined by the degree of fluctuations in the field. One is the constant term $\mathcal{G}_0$ given by $\frac{1}{\pi^2}\int r_1dr_1r_2dr_2 \langle \abs{\mathcal{E}(r_1)}^2\abs{\mathcal{E}(r_2)}^2 \rangle$. There are two terms whose contributions are equally important; one oscillates with a frequency $\ell_1\phi_0$ and the other with $\ell_2\phi_0$. The strength of these terms is determined by the quantity $\langle\Phi^{*}(r,0)\Phi(r,\phi_0)\rangle$, which is negligible for highly chaotic light ($r_0 \ll r$). The frequency of the fourth component is determined by the quantity $(\ell_1+\ell_2)\phi_0$, although its strength is dictated by the quantity $\langle\Phi^{*}(r,0)\Phi(r,\phi_0)\Phi^{*}(r,0)\Phi(r,\phi_0)\rangle$. For highly chaotic fields this is an extremely small contribution. The primary  contribution to $\langle I_{\ell_1}I_{\ell_2} \rangle$ is therefore given by the term
\begin{equation}
\begin{split}
\mathcal{G}_{\ell_1,\ell_2} =\frac{\alpha^4\text{sinc}^4 \left(\frac{[\ell_1-\ell_2]\alpha}{2}\right)}{4\pi^4} \int r_{1} dr_{1} r_{2} dr_{2}\abs{\mathcal{E}(r_{1})}^{2} \abs{\mathcal{E}(r_{2})}^{2}  (e^{-i(\ell_1-\ell_2)\phi_0}\langle \Phi^{*}(r_{1},0)\Phi(r_{1},\phi_{0})\Phi^{*}(r_{2},\phi_{0}) \Phi(r_{2},0) \rangle \\
+\text{c.c.})
\end{split}
\end{equation}
Note that this contribution describes an interference pattern that depends on the values of both $\ell_1$ and $\ell_2$.

\textbf{3. Orbital angular momentum correlations and angular position correlations. }

In this section we derive expressions for the correlations of pairs of OAM values and pairs of angular positions. The light that emerges from the DMD is given by $\mathcal{E}(r)\Phi(r,\phi)$. We make two copies of this field using a beam splitter and find the coincidences between projections onto two different modes of light. Let us first discuss the projection of one of the beams. The amplitude of the projection onto OAM mode $\ell$  is given by
\begin{align}
a_{\ell}=\int r dr d\phi \mathcal{E}(r)\Phi(r,\phi) \frac{e^{-i \ell\phi}}{\sqrt{2\pi}}g(r),
\end{align}
where $g(r)$ is the radial profile of the single-mode collection fiber, which is a Gaussian function. The intensity $|a_\ell|^2$   is given by
\begin{align}
I_{\ell}=\int r_1 dr_1 d\phi_1 \mathcal{E}(r_1)\Phi(r_1,\phi_1) \frac{e^{-i \ell\phi_1}}{\sqrt{2\pi}}g(r_1)
\times \int r_2 dr_2 d\phi_2 \mathcal{E^*}(r_2)\Phi^*(r_2,\phi_2) \frac{e^{i \ell\phi_2}}{\sqrt{2\pi}}g(r_2).
\end{align}
Therefore the ensemble-averaged intensity after the projection is given by 
\begin{align}
\langle I_{\ell}\rangle=\int r_1 dr_1 r_2 dr_2 d\phi_1 d\phi_2 \mathcal{E}(r_1)g(r_1) \mathcal{E^*}(r_2)g(r_2)  \frac{e^{-i \ell(\phi_1-\phi_2)}}{2\pi}\langle \Phi(r_1,\phi_1)\Phi^*(r_2,\phi_2) \rangle.
\end{align}
We are considering the case of highly fluctuating light;  in this regime
$\langle \Phi(r_1,\phi_1)\Phi^*(r_2,\phi_2) \rangle$ can be approximated by $ (1/r_1) \delta(r_2-r_1,\phi_2-\phi_1)$, leading to the result
\begin{align}
\langle I_{\ell}\rangle=\frac{1}{2\pi}\int r_1 dr_1  d\phi_1 \abs{\mathcal{E}(r_1)}^2g(r_1)^2.  
\end{align}
Note that this quantity is independent of the value $\ell$. Now let us consider the case of two coincident projections. The amplitude of coincident projections is given by 
\begin{align}
a_{\ell_1}a_{\ell_2}=\Pi^2_{i=1}\left(\int r_i dr_i d\phi_i \mathcal{E}(r_i)\Phi(r_i,\phi_i) \frac{e^{-i \ell_i\phi_i}}{\sqrt{2\pi}}g(r_i)\right).
\end{align}
We measure the rate at which these coincidences occur,  which is given by 
\begin{equation}
\begin{split}
I_{\ell_1}I_{\ell_2}=\int r_1 dr_1 d\phi_1 \mathcal{E}(r_1)\Phi(r_1,\phi_1) \frac{e^{-i \ell_1\phi_1}}{\sqrt{2\pi}}g(r_1)\times \int r_2 dr_2 d\phi_2 \mathcal{E}(r_2)\Phi(r_2,\phi_2) \frac{e^{-i \ell_2\phi_2}}{\sqrt{2\pi}}g(r_2)\\
\times \int r_3 dr_3 d\phi_3 \mathcal{E^*}(r_3)\Phi^*(r_3,\phi_3) \frac{e^{i \ell_1\phi_3}}{\sqrt{2\pi}}g(r_3)\times \int r_4 dr_4 d\phi_4 \mathcal{E^*}(r_4)\Phi^*(r_4,\phi_4) \frac{e^{i \ell_2\phi_4}}{\sqrt{2\pi}}g(r_4).
\end{split}
\end{equation}
After taking the statistical average, we obtain
\begin{equation}
\begin{split}
\langle I_{\ell_1}I_{\ell_2}\rangle&=\int r_1 dr_1 d\phi_1 \mathcal{E}(r_1)  \frac{e^{-i \ell_1\phi_1}}{\sqrt{2\pi}}g(r_1)\times \int r_2 dr_2 d\phi_2 \mathcal{E}(r_2)  \frac{e^{-i \ell_2\phi_2}}{\sqrt{2\pi}}g(r_2)\\&
\times \int r_3 dr_3 d\phi_3 \mathcal{E^*}(r_3)  \frac{e^{i \ell_1\phi_3}}{\sqrt{2\pi}}g(r_3)\times \int r_4 dr_4 d\phi_4 \mathcal{E^*}(r_4)  \frac{e^{i \ell_2\phi_4}}{\sqrt{2\pi}}g(r_4)
\times \langle \Phi(r_1,\phi_1)\Phi^*(r_3,\phi_3)\Phi(r_2,\phi_2) \Phi^*(r_4,\phi_4) \rangle.
\end{split}
\end{equation}
Following the same considerations and a similar procedure those used in obtaining Eq.\ \ref{eq:11},  we write the four-point coherence function as the sum of three contributions, each a product of two 2-point coherence functions. One of the contributions is always negligible for highly chaotic light. Another contribution leads to the simple product $\langle I_{\ell_1}\rangle \langle I_{\ell_2}\rangle$.  This contribution is actually independent of the values $\ell_1$ and $\ell_2$ in the limit of highly chaotic light for the same reason stated above in connection with  Eq.\ \ref{eq:9}.  The last contribution is given by
\begin{equation}
\begin{split}
&\int r_1 dr_1 d\phi_1 \mathcal{E}(r_1)  \frac{e^{-i \ell_1\phi_1}}{\sqrt{2\pi}}g(r_1)\times \int r_4 dr_4 d\phi_4 \mathcal{E^*}(r_4)  \frac{e^{i \ell_2\phi_4}}{\sqrt{2\pi}}g(r_4) \times \langle \Phi(r_1,\phi_1)\Phi^*(r_4,\phi_4)\rangle \\
&\times\int r_2 dr_2 d\phi_2 \mathcal{E}(r_2)  \frac{e^{-i \ell_2\phi_2}}{\sqrt{2\pi}}g(r_2)\times \int r_3 dr_3 d\phi_3 \mathcal{E^*}(r_3)  \frac{e^{i \ell_1\phi_3}}{\sqrt{2\pi}}g(r_3)\times 
 \langle\Phi(r_2,\phi_2) \Phi^*(r_3,\phi_3) \rangle.
 \end{split}
\end{equation}
By invoking the same approximation used above to evaluate the coherence functions as delta functions, we simplify this expression to
\begin{align}
\abs{\int r dr d\phi \abs{\mathcal{E}(r)}^2g(r)^2  \frac{e^{-i (\ell_1-\ell_2)\phi}}{2\pi}}^2. 
\end{align}
Note that the integral over $\phi$ vanishes unless $\ell_1=\ell_2$. Thus, we finally obtain
\begin{align}
\langle I_{\ell_1,\ell_2}\rangle=\langle I_{\ell_1}\rangle \langle I_{\ell_2}\rangle(1+\delta_{\ell_1,\ell_2}),
\end{align}
which is the expression used in the body of this paper to explain our experimental results.  Note that the correlations between two different values of OAM are half as large as those between the same value of OAM.

We can perform a similar calculation to find the correlations between two angular positions. For the case of a single beam (no beam splitter), the amplitude of the projection for a single value of $\phi$ is given by 
\begin{align}
a_{\phi}=\int r dr \mathcal{E}(r)\Phi(r,\phi) g(r).
\end{align}
It follows that the ensemble-averaged intensity at one of the detectors is given as follows
\begin{align}
\langle I_{\phi}\rangle &=  \langle |a_{\phi}|^2  \rangle \\&
 = \langle \int r _1 dr_1 \mathcal{E}(r_1)\Phi(r,_1\phi) g(r_1) \times \int r_2 dr_2 \mathcal{E}^*(r_2)\Phi^*(r_2,\phi) g(r_2)  \rangle  \\&
 = \int r _1 dr_1 r_2 dr_2  \mathcal{E}(r_1) \mathcal{E}^*(r_2) g(r_1) g(r_2) \,
 \langle \Phi(r,_1\phi) \Phi(r_2,\phi)^* \rangle
  \\& 
 = \int r dr \abs{\mathcal{E}(r)}^2 g(r)^2.
\end{align}
The last form follows from the substitution  $ \langle \Phi(r,_1\phi) \Phi(r_2,\phi)^* \rangle = (1/r_1) \delta(r_1 - r_2) $. 
If we now add the beam splitter and find the probability for coincidence detection of two beams, we obtain
\begin{align}
\langle I_{\phi_1}I_{\phi_2}\rangle&=\int r_1 dr_1  \mathcal{E}(r_1) g(r_1)\times \int r_2 dr_2  \mathcal{E}(r_2)  g(r_2)\\&
\times \int r_3 dr_3  \mathcal{E^*}(r_3) g(r_3)\times \int r_4 dr_4  \mathcal{E^*}(r_4)  g(r_4)
\times \langle \Phi(r_1,\phi_1)\Phi^*(r_3,\phi_3)\Phi(r_2,\phi_2) \Phi^*(r_4,\phi_4) \rangle.
\end{align}
Similar to the OAM case, we find that this expression yields two contributions. One of these contributions is independent of the values $\phi_1$ and $\phi_2$ whereas the other contribution is large only if the two intensities are measured at $\phi_1 = \phi_2$. This result can be described by the relation \begin{align}
\langle I_{\phi_1}I_{\phi_2}\rangle=\langle I_{\phi_1}\rangle \langle I_{\phi_2}\rangle [1+\delta(\phi_1-\phi_2)].
\end{align}
As expected, and similar to the case of correlating two OAMs, two non-overlapping angles share no correlation.  However, in contrast to the OAM variable, the angular position variable is not discrete, and one is allowed to correlate two regions defined by two angular positions that are not orthogonal, and consequently there is a partial overlap between the two correlated regions. Thus the degree of second-order correlation can take any value between 1 and 2.  Therefore an appropriate expression for this correlation function is given by

\begin{align}
\langle I_{\phi_1}I_{\phi_2}\rangle=\langle I_{\phi_1}\rangle \langle I_{\phi_2}\rangle[1+f(\phi_1-\phi_2)],
\end{align}
\noindent
where $f(\phi_1-\phi_2)$ is defined as 
\[
 f(\phi_1-\phi_2) =
  \begin{cases} 
      \hfill 0   \hfill &  \ \ \mbox{for} \ \   \abs{\phi_1-\phi_2}>\alpha/2\\
      \hfill 1-\frac{\abs{\phi_1-\phi_2}^2}{\alpha^2} \hfill &  \ \ \mbox{for} \ \  \abs{\phi_1-\phi_2}\leq\alpha/2. \\
  \end{cases}
\]
\noindent
$f(\phi_1-\phi_2)$ can be interpreted as the fractional  angular overlaps of the two slits.

\bibliography{Refns}{}
\bibliographystyle{naturemag.bst}

\vspace{4mm} \noindent\normalsize\textbf{Acknowledgements}
\vspace{4mm}

The authors would like to thank B. Rodenburg, S. Lukishova, J.J Sanchez-Mondragon, J. Wen, L. Gao, B. Gao, and Z. Yang  for helpful discussions. \textbf{Funding:} This work was supported by the U.S. Office of Naval Research, CONACyT, and the Mexican Secretaria de Edicacion Publica.  RWB acknowledges support from the Canada Excellence Research Chairs program.  \textbf{Author contributions:} O.S.M.-L. conceived the idea. O.S.M.-L. and M.M. designed the experiment. O.S.M.-L, R.M.C, M.M, and S.M.H.R. performed the experiment. R.M.C, S.M.H.R, O.S.M.-L, and M.M. performed the data analysis. The theoretical description was developed by S.M.H.R, O.S.M.-L, and M.M. The project was supervised by R.W.B. All authors contributed to the discussion of the results and to the writing of the manuscript.

\end{document}